\documentclass[letter, onecolumn]{IEEEtran}

\makeatletter
\def\ps@headings{%
\def\@oddhead{\mbox{}\scriptsize\rightmark \hfil \thepage}%
\def\@evenhead{\scriptsize\thepage \hfil\leftmark\mbox{}}%
\def\@oddfoot{}%
\def\@evenfoot{}}
\makeatother
\usepackage{cite}
\usepackage{amssymb,epsfig,latexsym,epsf,color}
\usepackage{graphicx,amsmath,amssymb,latexsym}
\usepackage{subfigure}
\usepackage{paralist}
\usepackage{bm}
\usepackage{setspace}
\usepackage{txfonts}

\newcommand{\sqeq}{\addtolength{\thinmuskip}{-4mu}
\addtolength{\medmuskip}{-4mu}\addtolength{\thickmuskip}{-4mu}}
\newcommand{\unsqeq}{\addtolength{\thinmuskip}{+4mu}
\addtolength{\medmuskip}{+4mu}\addtolength{\thickmuskip}{+4mu}}

\newcommand{\bea}{\begin{eqnarray}}
\newcommand{\eea}{\end{eqnarray}}
\newcommand{\beas}{\begin{eqnarray*}}
\newcommand{\eeas}{\end{eqnarray*}}
\newcommand{\bd}{\begin{displaymath}}
\newcommand{\ed}{\end{displaymath}}
\newcommand{\be}{\begin{equation}}
\newcommand{\ee}{\end{equation}}

\newcommand{\el}{\end{flushleft}}
\newcommand{\bl}{\begin{flushleft}}
\newcommand{\bc}{\begin{center}}
\newcommand{\ec}{\end{center}}
\newcommand{\remove}[1]{}

\newtheorem{theorem}{Theorem}

\newtheorem{lemma}{Lemma}

\def\cM{{\cal M}}

\def\cN{{\cal N}}

\def\cL{{\cal L}}

\def\and{\quad\mbox{and}\quad}

\def\bmg{{\bm{\gamma}}}

\def\bp{\noindent{\it Proof.}\ }
\def\ep{\hfill $\Box$}

\newcommand{\set}[1]{\mathcal{#1}}
\IEEEoverridecommandlockouts
\begin{document}

\title{Convergence and Tradeoff of Utility-Optimal CSMA}

\author{Jiaping Liu$^\dag$, Yung Yi$^*$, Alexandre Prouti\`ere$^\ddag$, Mung Chiang$^\dag$, and H. Vincent Poor$^\dag$\thanks{$\dag$:  Department of Electrical Engineering, Princeton University,
    USA. Email: \{jiapingl,chiangm,poor\}@princeton.edu. $*$: Department of Electrical Engineering and Computer Science, KAIST, South Korea. Email: {yiyung@ee.kaist.ac.kr}. $\ddag$: Microsoft Research, Cambridge, UK. Email: {alexandre.proutiere@microsoft.com.}  }}

\maketitle

\maketitle

\begin{abstract}
It has been recently suggested that in wireless networks, CSMA-based distributed MAC algorithms could achieve optimal utility without any message passing. We present the first proof of convergence of such adaptive CSMA algorithms towards an arbitrarily tight approximation of utility-optimizing schedule. We also briefly discuss the tradeoff between optimality at equilibrium and short-term fairness practically achieved by such algorithms.  

\end{abstract}

\section{Model and Algorithm}

\subsection{Network model and optimization problem}

Consider a wireless network composed by a set $\cL$ of $L$
interfering links. Interference is modeled by a symmetric, boolean matrix $A\in
\{0,1\}^{L\times L}$, where $A_{lk}=1$ link $l$
interferes link $k$. 
Define by $\cN \subset \{0,1\}^L$ the set of feasible link activation profiles, or,
schedules. A schedule $m\in \cN$ is a subset of
non-interfering active links. 
The transmitter of link $l$ can transmit
at a fixed unit rate when active, and all links are saturated with infinite backlog.

Denote
by $\bmg=(\gamma_l,l\in \cL)$ the long-term throughputs achieved by a scheduling algorithm, which determines which links are activated at each time. Let
$U:\mathbb{R}^+\to \mathbb{R}$ be an increasing and strictly concave
objective function. The following utility optimization problem over schedules has been extensively studied: 
\begin{align}\label{eq:ut1}
\max \quad  & \Sigma_{l\in \cL}U(\gamma_l),\quad \hbox{s.t. }\forall l\in {\cal L}, \gamma_l \le \Sigma_{m:l\in m}\pi_m\\
& \forall m\in \cN, \pi_m\ge 0, \Sigma_{m\in \cN}\pi_m = 1\nonumber.
\end{align}
where $\pi_m$ is the long-term proportion of time when schedule $m$ is
used. We denote by $\overline{{\bm{\gamma}}}=(\overline{\gamma}_l,l\in
\cL)$ the solution of (\ref{eq:ut1}).  Most of the proposed distributed
schemes to solve (1) make use of a dual decomposition of the problem into a rate
control and a scheduling problem: A
virtual queue is associated to each link; a rate control algorithm
defines the rate at which packets are sent to the virtual queues, and a
scheduling algorithm decides, depending on the level of the virtual
queues, which schedule to use with the aim of stabilizing all virtual
queues. The main challenge reduces to developing a distributed
and efficient scheduling algorithm. Most of the proposed solutions are
semi-distributed implementations of the max-weight scheduler introduced in \cite{TE92}, and require
information about the queues to be passed around among the nodes or links
(e.g., see a large set of references in \cite{YPC08}).
This signaling overhead increases communication complexity and 
reduces effective throughput of the algorithms.  

Recently, there have been proposals that do not use any
message passing, and yet achieve high efficiency \cite{walrand,RS08,techrep}. These algorithms are based on the random access protocol of 
CSMA (Carier Sense Multiple Access), and leverage simulated annealing techniques to solve the max-weight scheduling problem as first proposed in \cite{BH88}.  
In this paper we provide the first rigorous proof that such adaptive CSMA algorithms indeed converge, 
using stochastic approximation tools with two time-scales.
Then, we quantify the impact of inevitable collisions in CSMA
protocols on the trade-off between their long-term efficiency and short-term 
fairness, the latter defined here as $\beta={1 / \max\{E_l, l\in \cL\}}$, where $E_l$ is the average duration during which link $l$ do not transmit successfully. It turns out that a price to pay for the surprisingly good utility performance by these message-passing-free algorithms  is short-term starvation. 

\subsection{Adaptive CSMA algorithms}

To access the channel, each transmitter $l \in \set{L}$ runs a random back-off algorithm
parametrized by two positive real numbers $(\lambda_l,\mu_l),$ denoted as 
CSMA$(\lambda_l,\mu_l)$: after a successful
transmission, the transmitter randomly picks a back-off counter
according to some distribution of mean $\lambda_l$; it decrements the
counter only when the channel is sensed idle; and it starts transmitting
when the back-off reaches 0, and remains active for a duration $\mu_l$.

We first consider a continuous-time and then a discrete-time
version of CSMA algorithms. In the ideal continuous-time setting (considered in this and the next sections), the
back-off counter distribution is exponential, so that two interfering
links cannot be activated simultaneously and collisions are avoided. In
practice, back-offs are discrete, say geometrically distributed. In this discrete setting (studied in Section III), collisions may occur, 
degrade performance, and introduce tradeoff between long-term utility and short-term fairness. 

If parameters $(\lambda_l,\mu_l), l\in {\cal L}$ were fixed, the analysis of the dynamics of systems under {\it continuous-time} CSMA algorithms would be classical (e.g., \cite{DT06} and references therein). In this case, in steady state the set of active links evolves according to a reversible Markov process whose stationary distribution, denoted by $\pi^{\bm{\lambda},\bm{\mu}}$, is defined by:
\begin{equation}\label{eq:st}
\forall m\in {\cal N}, \pi^{\bm{\lambda},\bm{\mu}}(m)={\prod_{l\in m}\lambda_l \mu_l\over \sum_{m'\in {\cal N}}\prod_{l\in m'}\lambda_l\mu_l}. 
\end{equation}

We now describe how transmitters adapt their CSMA parameters. Time is slotted and transmitters update their parameters at the beginning of each slot. To do so, they maintain a virtual queue, denoted by $q_l[t]$ in slot $t$, for link $l$. The algorithm operates as follows (the algorithm presented here is an extension of those proposed in \cite{walrand}):\\
\textbf{Algorithm 1.}
\begin{enumerate}
\item During slot $t$, the transmitter of link $l$ runs CSMA($\lambda_l[t],\mu$), and records the amount $S_l[t]$ of service received during this slot;
\item At the end of slot $t$, it updates its virtual queue and its CSMA parameters according to:
\begin{equation*}
\label{eq:q_alg1}
q_l[t+1]=\bigg[q_l[t]+{b[t]\over W'(q_l[t])}\big(U'^{-1}({W(q_l[t])\over
  V})-S_l[t]\big)\bigg]^{q^{\max}}_{q^{\min}}, 
\end{equation*}
and set
$\lambda_l[t+1]=\mu^{-1}\exp \{ W(q_l[t+1])\}$.
\end{enumerate}

At the beginning of each slot, each non-active transmitter picks a new random back-off counter to account for the CSMA parameter update.
In Algorithm 1, $b:\mathbb{N}\to \mathbb{R}$ is a step size
function; $W:\mathbb{R}^+\to \mathbb{R}^+$ is a strictly increasing and
continuously differentiable function, termed {\it weight function}; 
$V$, $q^{\min}$, $q^{\max} (>q^{\min})$ are positive
parameters, and $[\cdot]_c^d\equiv \max(d,\min(c,\cdot))$.

\section{Convergence Proof}

The main difficulty in analyzing the convergence of Algorithm 1 lies in the
fact that the updates in the virtual queues depend on random processes
$(S_l[t], t\ge 0)$, whose transition rates in turn depend on the virtual
queues. As we will demonstrate, it is possible to represent Algorithm 1
as a stochastic approximation algorithm with controlled Markov noise as
introduced by Borkar \cite{borkar06}.  

We use the notation $\pi^{\bm{q}}$ to denote the distribution on $\cN$ defined by:
\begin{equation}\label{eq:pi_q}
\forall m\in \cN, \quad \pi^{\bm{q}}(m)={\exp (\sum_{l\in m}W(q_l))\over \sum_{m'\in {\cal N}}\exp(\sum_{l\in m'}W(q_l))}.
\end{equation}
We also denote by $\bm{\gamma}[t]=(\gamma_l[t]={1\over
  t}\sum_{i=0}^{t-1}S_l[i], l\in \cL)$ the random variable representing
the time-averaged service rates received by the various links over the
interval $[0,t).$ The following theorem states the convergence of Algorithm 1, for diminishing step-sizes (similar but weaker results are readily obtained for constant, small step-sizes).

\medskip
\begin{theorem}
\label{theo:asym_PF}
Assume that $V\le W(q^{\max})/U'(1)$ and that $\sum_{t=0}^\infty b[t]=\infty,\quad \sum_{t=0}^\infty b[t]^2<\infty$. For any initial condition $\bm{q}[0]$, under Algorithm 1, we have the following convergence:
$$
\lim_{t\to\infty}{\bm{q}}[t]=\bm{q}^{\star} \textrm{ and } \lim_{t\to \infty}\bm{\gamma}[t]=\bm{\gamma}^{\star}, \textrm{ almost surely},
$$
where $\bm{q}^{\star}$ and $\bm{\gamma}^{\star}$ are such that: $(\bm{\gamma}^{\star},\pi^{\bm{q}^{\star}})$ solves:
\begin{eqnarray}\label{eq:OPT}
\mbox{max} ~&&V\Sigma_{l\in \cL} U(\gamma_l)-\Sigma_{m\in\cN} \pi_m\log \pi_m\cr
\mbox{s.t.}~ &&\gamma_l\leq \Sigma_{m\in \cN:m_l=1} \pi_m, \quad\Sigma_{m\in \cN} \pi_m=1.
\end{eqnarray}
Furthermore, Algorithm 1 approximately solves (\ref{eq:ut1}) as:
\begin{equation}
\label{eq:gap}
\big| \Sigma_{l\in \cL}\big(U(\overline{\gamma}_l)-U(\gamma_l^{\star})\big) \big| \le {\log\vert \cN\vert/V}.
\end{equation}
\end{theorem}

\medskip \bp We first
show that the network dynamics under the continuous-time random
CSMA protocol can indeed be averaged and it asymptotically approaches to a
deterministic trajectory (see Lemma~\ref{lem:nf_conv}). Resolving this bottleneck in understanding adaptive CSMA is the main contribution in this proof. Then we prove that the resulting averaged
algorithm converges to the solution of (\ref{eq:OPT}).

\noindent
\underline{Step 1.} From the discrete-time sequence $(\bm{q}[t],t\ge 0)$, we define a continuous function $\bar{\bm{q}}(\cdot)$ as follows. Define for all $n$, $t_n=\sum_{i=1}^nb[i]$, and for all for all $t_n<t\le t_{n+1}$,
\begin{equation}
\label{eq:q_bar}
\bar{q}_l(t)=q_l[n]+(q_l[n+1]-q_l[n])\times({t-t_n\over t_{n+1}-t_n}).
\end{equation}

\begin{lemma}[Convergence and averaging]
\label{lem:nf_conv}
Denote by $\tilde{q}$ the solution of the following system of o.d.e.'s:
for all $l$,
\sqeq
\begin{equation}
\label{eq:q_ode}
{d\tilde{q}_l / dt}=\left(U'^{-1}\Big({W_l(\tilde{q}_l)/ V}\Big)-\Sigma_{m\in \cN:m_l=1} \pi^{\tilde{q}}(m)\right)\cdot {\bm{1}_{\{ q^{\min} \leq \tilde{q}_l\leq q^{\max}\}}\over W'(\tilde{q}_l)},
\end{equation}
\unsqeq
with $\tilde{\bm{q}}(\tau)=\bar{\bm{q}}(\tau)$. Then we have: for all $T>0$,
\begin{equation}
  \lim_{\tau\rightarrow \infty}\sup_{t\in [\tau, \tau+T]} \|\bar{\bm{q}}(t)-\tilde{\bm{q}}(t)\|=0 \quad \text{a.s.}
\end{equation}
\end{lemma}

Lemma~\ref{lem:nf_conv} shows that the trajectory of the continuous
interpolation $\bar{{\bm{q}}}$ of the sequence of the virtual queues
${\bm{q}}$ asymptotically approaches that of $\tilde{\bm{q}}$. Note that in the limiting o.d.e.'s, the service $S_l$
received on each link is averaged (as if the virtual queues were
frozen), and this averaging property constitutes the key challenge in
analyzing the convergence of Algorithm 1. 

To prove
Lemma~\ref{lem:nf_conv},  we attach to each link $l$ a variable
$a_l[t],$ where $a_l[t]=1$ if the link is active at time $t$ (at the
end of slot $t$), and $0$ otherwise. Now it can be easily seen
that $\bm{Y}[t]=(\bm{S}[t],\bm{a}[t])$ is a non-homogeneous Markov chain
whose transition kernel between times $t$ and $t+1$ depends on
$\bm{q}[t]$ only (this can be checked as in \cite{DT06}). Now the
updates of the virtual queues in Algorithm 1 can be written as:
$$
q_l[t+1]=q_l[t]+b[t]\times h(q_l[t],Y_l[t]),
$$
where
$$
h(q,Y)=\frac{1}{W'(q)}(U'^{-1}(W(q)/V)-S_l))\cdot \bm{1}_{ \{q^{\min}\leq q_l\leq q^{\max}\}}.
$$
As a consequence, Algorithm 1 can be seen as a stochastic approximation
algorithm with controlled Markov noise as defined in \cite{borkar06,borkar}. To complete the proof of Lemma 1, we check the
conditions as stated in \cite{borkar}.\\
1) The transition kernel of $\bm{Y}[t]$, parametrized by $\bm{q}[t]$, is continuous in $\bm{q}[t]$ (because the transition rates from one state to another are determined by the $\lambda_l[t]$'s, which are continuous in the $q_l[t]$'s). Note also that fixing $\bm{q}[t]=\bm{q}_0$, the obtained Markov chain $\bm{Y}[t]$ is ergodic (its state-space is finite) with stationary distribution $\pi^{\bm{q}_0}$.\\
2) $h$ is continuous and Lipschitz in the first argument, uniformly in the second argument. This can be easily checked, given the properties of the utility and weight functions $U$ and $W$ and observing that we restrict our attention to the compact $[q^{\min},q^{\max}]$.\\
3) {\it Stability} condition: $q_l[t]<q^{\max}$ for all $l\in \cL$ and $t\ge 0$.\\
4) {\it Tightness} condition (corresponding to ($\dagger$) in \cite{borkar}[pp.71]): This is satisfied since $\bm{Y}[t]$ has a finite state-space (cf. conditions (6.4.1) and (6.4.2) in \cite{borkar}[pp.76]). 

This completes the proof of Lemma~\ref{lem:nf_conv}.  By Lemma 1, if
there exists an equilibrium $\bm{q}^{\star}$ such that
$\lim_{t\to\infty}\tilde{\bm{q}}(t)=\bm{q}^{\star}$, then we would also
have: $\lim_{t\to\infty}{\bm{q}}[t]=\bm{q}^{\star}$ a.s. (see
\cite{benaim96} for details).

\noindent
\underline{Step 2.} To complete the proof, we show that
(\ref{eq:q_ode}) may be interpreted as a sub-gradient algorithm
(projected on a bounded interval) solving the Lagrange dual
problem of (\ref{eq:OPT}). Let $D(\bm{\nu}, \eta)$ denote the dual
function of (\ref{eq:OPT}). Then we show that (\ref{eq:q_ode}) is the
sub-gradient algorithm of:
\begin{eqnarray}
\label{eq:OPT_dual}
\mbox{min} ~ D(\bm{\nu}, \eta), ~\mbox{s.t.} ~\nu^{\min} \leq \nu_l\leq \nu^{\max}, ~\forall l\in \cL.
\end{eqnarray}
Here we include the upper-bound $\nu^{\max}$ (resp. lower-bound
$\nu^{\min}$) that corresponds to the limitation of the $q_l$'s:
$\nu^{\max}=W(q^{\max})$ (resp. $\nu^{\min}=W(q^{\min})$).
The Lagrangian of (\ref{eq:OPT}) is given by:
\begin{align*}
L(\bm{\gamma}, \bm{\pi};\bm{\nu},\eta)=&\big(\Sigma_{l\in \cL} V\log \gamma_l-\nu_l\gamma_l\big) +\big(\Sigma_{l\in \cL} \nu_l\Sigma_{m\in \cN: m_l=1} \pi_m \\
& -\Sigma_{m\in \cN} \pi_m\log \pi_m\big)+\eta\big(\Sigma_{m\in \cN} \pi_m-1\big).
\end{align*}
Then the KKT conditions of (\ref{eq:OPT}) are given by:
\begin{eqnarray}
  \label{eq:KKT_gamma} &&VU'(\gamma_l)=\nu_l, ~\forall l\in \cL,\\
  \label{eq:KKT_pi}&&-1-\log \pi_m+\Sigma_{l:m_l=1} \nu_l+\eta=0, ~\forall m\in \cN,\\
  \label{eq:KKT_nu} && \nu_l\big(\gamma_l-\Sigma_{m\in \cN:m_l=1} \pi_m\big)=0.
\end{eqnarray}
Now if $\nu_l=W(\tilde{q}_l)$ for all $l$, (\ref{eq:KKT_pi}) is solved
for $\pi^{\tilde{\bm{q}}}$ (in view of (\ref{eq:pi_q})). The
sub-gradient of (\ref{eq:KKT_nu}) (when accounting for
(\ref{eq:KKT_gamma})) is:
\begin{equation}\label{eq:ode2}
  {d\nu_l / dt}=\left(U'^{-1}\Big({\nu_l / V}\Big)-\Sigma_{m\in
      \cM \atop :m_l=1} \pi_m^{\tilde{\bm{q}}}\right)\cdot \bm{1}_{\{ \nu^{\min}\leq \nu_l\leq \nu^{\max}\}},
\end{equation}
which is equivalent to (\ref{eq:q_ode}), provided that the solution
$\nu_l^{\star}$, $l\in {\cal L}$, of (\ref{eq:OPT_dual}) without the
constraints $\nu^{\min}\le \nu\le\nu^{\max}$ actually belongs to the
interval $[\nu^{\min},\nu^{\max}]$. The latter condition is satisfied by
simply combining $\gamma_l\le 1$, (\ref{eq:KKT_gamma}), and the
assumption $V\le \nu^{\max}/U'(1)$.  Finally, since (\ref{eq:OPT_dual})
is a strictly convex optimization problem, (\ref{eq:ode2}) converges to its unique equilibrium
$\bm{\nu}^{\star}$. Finally, the inequality (\ref{eq:gap}) is obtained comparing (\ref{eq:ut1}) and (\ref{eq:OPT}), because entropy $-\sum_m\pi_m\log \pi_m$ is always bounded by $\log \vert {\cal N}\vert$. The proof of Theorem~\ref{theo:asym_PF} is
completed. \ep

\section{Short-term fairness vs. long-term efficiency}

In the previous section, we have analyzed the convergence of Algorithm 1 in the ideal, continuous-time setting of CSMA protocols with no collisions. In practical implementaions however, the back-off counters are discrete and collisions possible \cite{techrep,srikant}. To keep collision rates less than $\epsilon$, we scale down in Algorithm 1 the transmission probabilities to $\epsilon\lambda_l[t]$ and scale up the channel holding time to $\mu/\epsilon$. Denote by ${\bm{\gamma}}^{\star}_{\epsilon}$ the throughput vector obtained with this modified algorithm. Of course, we have $\lim_{\epsilon\to 0} {\bm{\gamma}}^{\star}_{\epsilon}={\bm{\gamma}}^{\star}$. More precisely, using standard perturbation analysis, one can show that for all $l$:
$$
{\gamma}^{\star}_{\epsilon,l}=\gamma_l^{\star} - C_l\epsilon+o(\epsilon).
$$
The constants $C_l$ can be derived explicitly for networks with simple interference structures, but are more difficult to obtain for general networks. For example, in networks with full interference, i.e., where all links interfere each other, we can easily prove that $C_l$ is roughly equal to $\lambda_l^{\star}$, which in turn is equal to $\lambda^{\max}=\mu^{-1}\exp{W(q^{\max})}$ (indeed, in view of (\ref{eq:st}), the throughput on any link is just equal to $\lambda_l^{\star}\mu/(1+L\lambda_l^{\star}\mu)$, an increasing function of $\lambda^{\star}_l$). Combining the above observation and (\ref{eq:gap}), the distance between ${\bm{\gamma}}^{\star}_{\epsilon}$ and the utility optimal vector  $\bar{\bm{\gamma}}$, which represents the efficiency gap of the algorithm, scales as $k_1/V  + \epsilon k_2({\bm{\lambda}}^{\star})$ when $\epsilon$ is small, where $k_1$ is a positive constant, and $k_{2}$ a constant that depends on ${\bm{\lambda}}^{\star}$). From the assumption on $V$ in Theorem 1, we deduce that the efficiency gap scales as $k_1/\log(\lambda^{\max}\mu) + \epsilon k_2({\bm{\lambda}}^{\star})$. For network with full interference, the gap scales as: ${k_1/ \log(\lambda^{\max}\mu)} + \epsilon \lambda^{\max}$.

Let us now evaluate the short-term fairness index. Using cycle formula
\cite{babre01}, at the equilibrium, the average of periods, during which link $l$ do not transmit successfully, is given by:
$$
E_l = {\mu\over \epsilon} \times {1-\gamma_{\epsilon,l}^{\star}\over \gamma_{\epsilon,l}^{\star}}={\mu\over \epsilon} \times \left({1-\gamma_{l}^{\star}\over \gamma_{l}^{\star}}+o(\epsilon)\right).
$$
Then the short-term fairness index $\beta$ scales as $k_3\cdot\epsilon /\mu$, where $k_3$ is a positive constant. 

Now if we want to guarantee an efficiency gap less than $\delta$, in view of the above analysis, we must have $\lambda^{\max} \mu \ge \exp(k_1/\delta)$. In the case networks with full interference, this further implies that:
$\beta \le {k_3 \delta / \exp(k_1/\delta)}$. For networks with more general interference structure, analytically expressing the tradeoff between efficiency and short-term fairness is more difficult, but the conclusion remains similar: to approach optimality in sum utility in the long-run, some node will be denied channel access for longer time, and we must pay a price of short-term unfairness that grows very fast (the channel holding time must grow like $e^{1/\delta}/\delta$).

\begin{figure}[t]
\centering
\includegraphics[width=0.5\columnwidth]{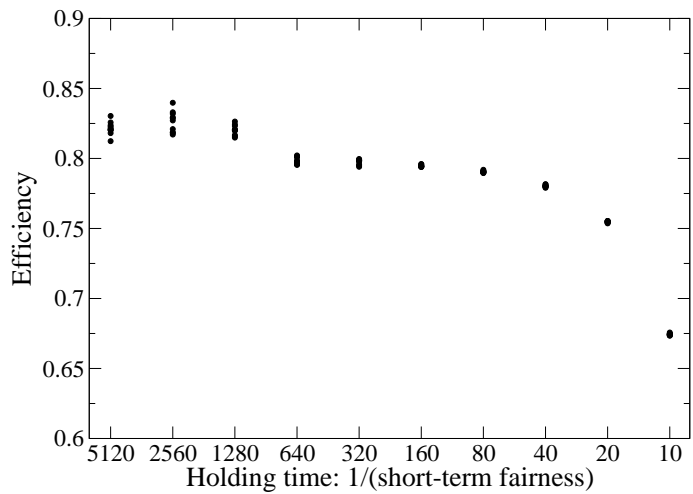}
\caption{ \label{fig:tradeoff}Efficiency vs. short-term fairness tradeoff in a 3-link linear network. Algorithm parameters: $b[t]=0.001$, $W(x)=x$, $V=1$, $\epsilon\lambda^{\max}=0.1$.}
\end{figure}

We illustrate this tradeoff numerically on a simple 3-link linear network, in which link 1 and link 3 both interfere with link 2 but link 1 and link 3 do not interfere. Figure~\ref{fig:tradeoff} shows the efficiency (i.e., $1-\delta$) as a function of 1/(short-term fairness index). 10 experiments were carried out with different random seeds for each value in $x$-axis. 
In the practical setting with collisions, 85\% efficiency, in terms of utility achieved, is quite good for random access without message passing, although this efficiency drops as short-term fairness improves exponentially. 

\bibliographystyle{IEEEtran}

\end{document}